\documentclass[a4paper,11pt]{article}
\usepackage{jinstpub} 
\usepackage{lineno}
\usepackage[flushleft]{threeparttable}
\usepackage{float}
\usepackage{caption}
\usepackage{subfigure}
\usepackage{graphicx}
\graphicspath{ {figs/} }
\usepackage{multirow}
\usepackage[euler]{textgreek}
\usepackage[dvipsnames]{xcolor}
\usepackage{ulem}
\usepackage{soul,xcolor}
\title{\boldmath Designing the system to measure the depth-dose profile of a proton beam using CsI(Tl) scintillator}
\author[a,1]{P.~Boontueng,\note{Corresponding author.}}
\author[b]{N.~Ritjoho,}
\author[b]{P.~Phumara,}
\author[c]{T.~Sanghangthum,}
\author[d]{P.~Thongjerm,}
\author[d]{S.~Wonglee,}
\author[d]{W.~Pornroongruengchok,}
\author[b]{and C.~Kobdaj}
\affiliation[a]{Department of Physics, Faculty of Science, Ubon Ratchathani University (UBU), Ubon Ratchathani 34190, Thailand}
\affiliation[b]{School of Physics, Suranaree University of Technology (SUT), Nakhon Ratchasima, 30000 Thailand}
\affiliation[c]{Division of Radiation Oncology, Department of Radiology, Faculty of Medicine, Chulalongkorn University (CU), Bangkok 10330, Thailand}
\affiliation[d]{Thailand Institute of Nuclear Technology (TINT), Nakhon Nayok 26120, Thailand}
\emailAdd{phongnared.b@ubu.ac.th }
\abstract{In this work, the standard CsI(Tl) scintillator was used to determine the characteristics of a proton beam. By irradiating the scintillator with a proton beam, it was able to subsequently measure the emitted light using a spectrometer. This work presents the dose response of a scintillator and its use to measure the depth-dose profile of a proton beam. The measurement of the dose response showed that the emission light of the scintillator depended on the total dose of the proton beam. For the design of the depth-dose measurement, the setup of the water phantom included a scintillator and a water tank. The depth-dose profile was determined through varying the water depth at various positions along the path of the proton beam until the end of the Bragg peak region. To compare the measured depth-dose profile, the deposited energy was simulated and fitted Birks’ constant for collecting quenching. Furthermore, the recommended measurement design suggested using a thin scintillator, as it resulted in a narrower shape of the depth-dose profile. Based on experiment and simulation data, a standard CsI(Tl) scintillator was promising for determining the characteristic proton beam.}


\begin{document}
\maketitle
\flushbottom
\section{Introduction}
\label{sec:intro}
\indent The scintillating materials have been widely used as radiation detectors as well as survey meters. They have become an important part of a radiation dosimeter used for treatment planning to measure the total dose. Proton therapy is an alternative therapy technique to treat cancer patients due to its ability to deliver a high dose at the end of its traveling path, called the Bragg peak \cite{ref1}. The depth of protons in a patient depends on the initial proton energy, which can be dominated in the region of cancer cells for proton therapy. The depth dose of a proton beam can be determined by measuring the stopping power ($dE/dx$) using the scintillating materials through a light emission. During the irradiating proton beam to the scintillator, protons ionize electrons in a ground state, excite them to a higher state, and subsequently release their energy to the luminescent center. After that, excited electrons released their energy into the ground state in the form of a light emission. This is known as the scintillation process. Many scintillating materials were proposed to measure the depth-dose profile, e.g., plastic scintillators \cite{ref2,ref3,ref4,ref5,ref6,ref7}, bare silica fibers \cite{ref8}, Gd$^{3+}$-doped silica fibers \cite{ref9}, and Ce$^{3+}$-doped glasses \cite{ref9,ref10,ref11,ref12}. 
\newline \indent The linear energy transfer (LET) of the proton deposited into matter significantly changes along a proton path. In the region of the Bragg peak, the amount of light emission should be proportional to the deposition energy. However, the light emission of most scintillating materials was suppressed in the high LET region \cite{ref2,ref3,ref4,ref5,ref6,ref7,ref8,ref9,ref10,ref11,ref12,ref13,ref14}. This phenomenon was known as quenching, as described by Birks’ Law \cite{ref15}. There were many methods to measure the profile of the proton beam using the scintillator. The measurements of the depth-dose profile for the proton beam were presented by using the scintillator with a thickness that covered the proton range \cite{ref2}. However, the scintillator in the range of 250 µm diameter or less was required to measure the depth-dose by irradiating a proton beam to the scintillator by changing the water depth \cite{ref8,ref9,ref11,ref14}. 
\newline \indent In this work, we studied the response of the scintillator to distinguish the different doses of the proton beam. Then, a standard CsI(Tl) scintillator was used to determine the depth-dose of the proton beam by varying the water depth. We used the GATE packages (version 9.2), a Monte Carlo simulation programming, to simulate the deposited energy of the proton beam \cite{ref16,ref17,ref18}. The primary protons were demonstrated with 10$^6$ particles with the physics list QGSP\_BIC\_EMY, which is appropriate for hadrons \cite{ref19}. The Birks’ constant for the scintillator was calculated and proposed to fit the depth-dose profile. Furthermore, a simulation of the recommended setup design for measuring the depth-dose profile was presented.
\section{Materials and methods}
\subsection{Materials and the light measurement}
\label{sec:2.1}
\indent The standard CsI(Tl) scintillator was obtained from Kinheng Crystal Material (Shanghai) Co., Ltd. Its dimension was 15 × 10 × 3 mm$^3$, with a density of 4.51 g/cm$^3$. The light emission from the scintillator under proton irradiation was detected by the 543-nm Air-Spaced Doublet Collimator (F810SMA-543), and then the light was sent to the spectrometer (AvaSpec-ULS4096CL-EVO) by a SMA fiber optic. The spectrometer collected the light emission with an integration time of 6 seconds to cover the irradiating time. The signal was recorded and analyzed by AvaSoft (version 8.14.0.0). The dark spectrum (background light) was from output signals to correct the raw spectrum in mode “Scope minus Dark Mode.”\\

\subsection{Dose response}
\label{sec:2.2}
\indent To investigate the response of the standard CsI(Tl) scintillator to a proton beam, the scintillator was irradiated with a 70 MeV proton beam and a current of 300 nA at various total doses. The light yield per proton dose was used to evaluate the scintillator's response. The proton beam, generated by the Varian ProBeam Compact Therapy System at the King Chulalongkorn Memorial Hospital (KCMH), Thailand, was prepared through a quality assurance (QA) procedure that ensured the desired energy and current. A 3 cm-thick virtual water phantom (VWP) with a composition of H:7.7$\%$, C:68.7$\%$, N:2.3$\%$, O:18.9$\%$, Cl:0.1$\%$, and Ca:2.3$\%$ (a density of 1.03 g/cm$^3$) was placed in front of the scintillator to shift the Bragg peak into the scintillator \cite{ref10}. In this study, the scintillator was exposed to proton doses of 10, 17.5, and 25 Gy. The measurement system was kept in the dark box to prevent background light. The scintillator was positioned at the isocenter of the gantry.\\

\subsection{Depth-dose profile measurement}
\begin{figure}[htp] 
\centering
\includegraphics[width=0.7\textwidth]{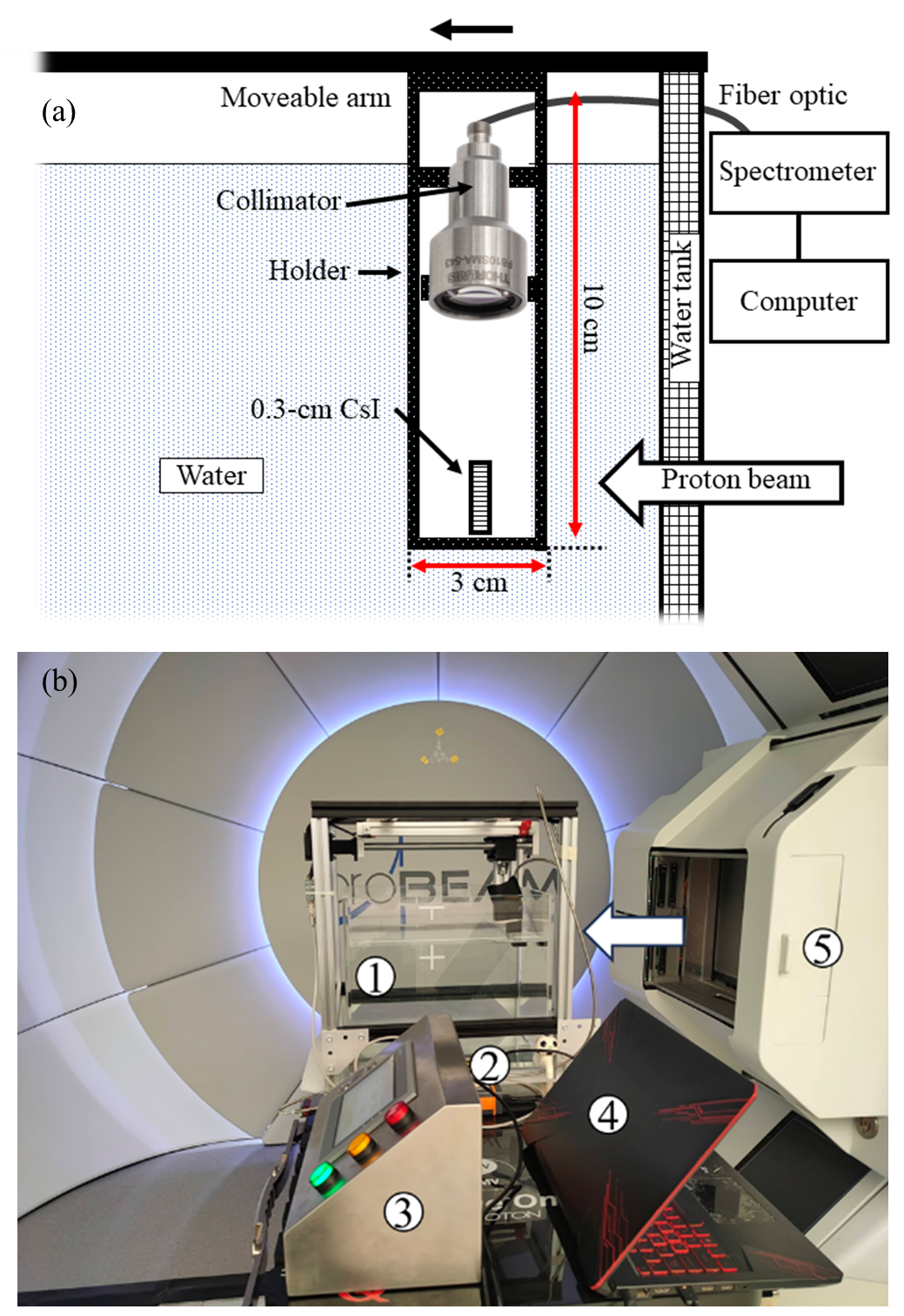}
\caption{(a) The schematic setup of the water phantom to determine the depth-dose profile of the proton beam. (b) The measurement was set up in the treatment room at KCMH, where there were (1) a water phantom, (2) a spectrometer, (3) a water-depth controller, (4) a computer and data storage, and (5) a proton nozzle. The direction of the proton beam is represented by the white arrow.}
\label{fig:1}%
\end{figure}

\indent A depth-dose profile of the proton beam was obtained by measuring light output at different depths within a water phantom, as presented in \cite{ref11}. Figure \ref{fig:1}(a) shows the schematic setup of the depth-dose profile using a scintillator in the water phantom. The measurement system consisted of a moveable arm, a light measurement system, and a water tank. A 3 mm-thick scintillator was placed within a holder and attached to the moveable arm, allowing for precise positioning within the water tank where the wall thickness was 1 cm. The water level above the scintillator was adjusted, with a minimum depth of 0.5 cm. The proton beam was directed at the center of the scintillator, as indicated by the white arrow in figure \ref{fig:1}(a). Figure \ref{fig:1}(b) presents the measurement system placed on top of the patient bed, which includes the water phantom, the light measurement system, the water-depth controller, and the data acquisition components. The proton beam was shot horizontally toward the scintillator. The scintillator was exposed to a 70-MeV proton beam at a fixed dose of 25 Gy and a beam current of 300 nA, which was the same as in Section \ref{sec:2.2}. Finally, the measured data were compared with the simulation results.\\

\subsection{Quenching correlation factor}
\indent The correlation of the light emission from the scintillator was fitted by the first-order quenching of Birks’ law \cite{ref15}, given by equation \ref{eq:brik},
\begin{equation}
\label{eq:brik}
    \frac{dL}{dx}=\frac{S\frac{dE}{dx}}{1+k_{B}\frac{dE}{dx}}
\end{equation}
where $L$ represents the emitted light from the scintillator, $S$ is the scintillation efficiency, $dE/dx$ is the proton stopping power (which $E$ represented the deposited energy resulting from the simulations), and $k_B$ was the Birks’ constant.

\section{Results and discussions}
\begin{figure}[htp] 
\centering
\includegraphics[width=0.8\textwidth]{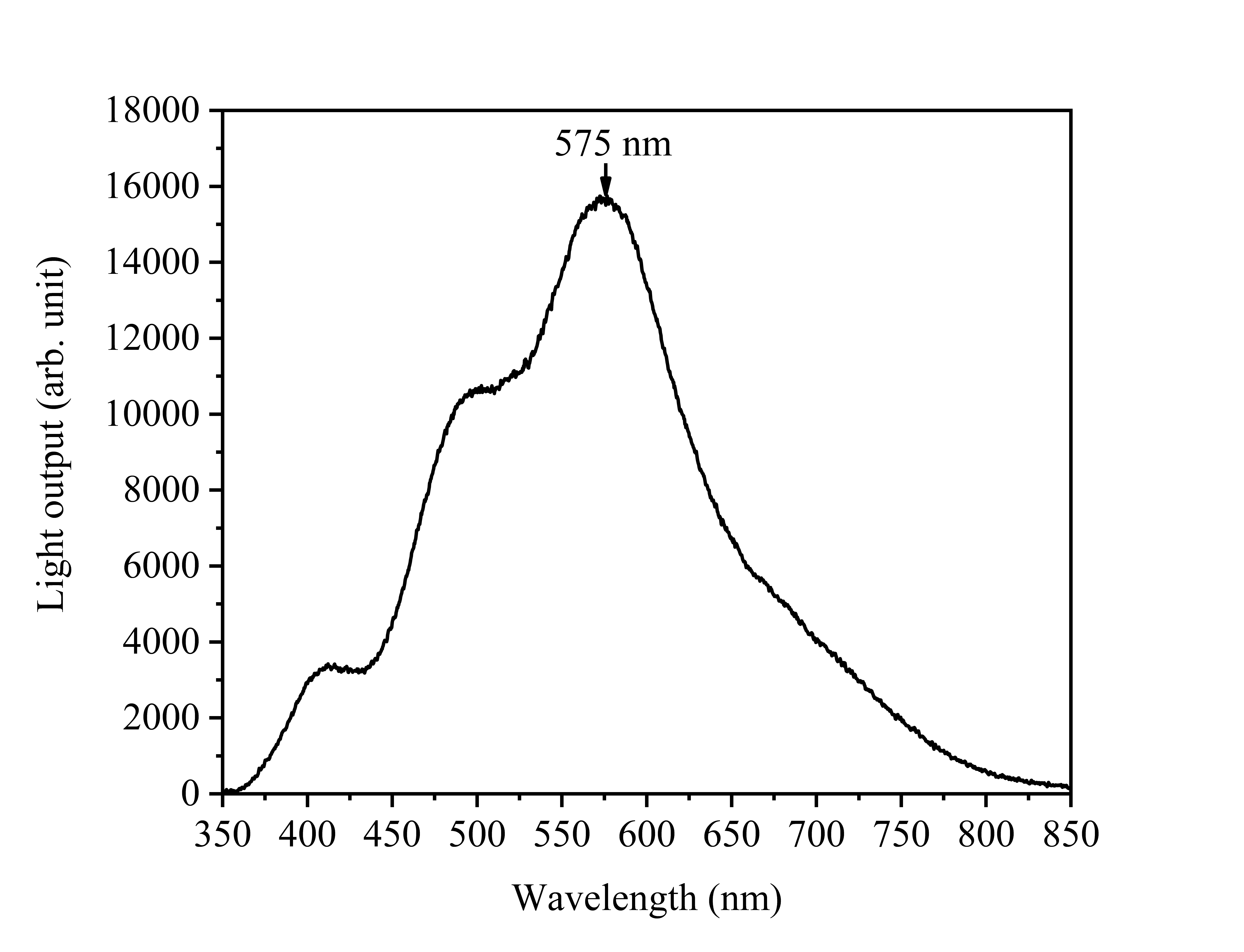}
\caption{The light emission wavelength of the standard CsI(Tl) scintillator, which was irradiated by the 70 MeV proton with a total dose of 25 Gy in the setup of the 3 cm-thick VWP as a range shifter.}
\label{fig:2}%
\end{figure}

\begin{figure}[htp] 
\centering
\includegraphics[width=0.8\textwidth]{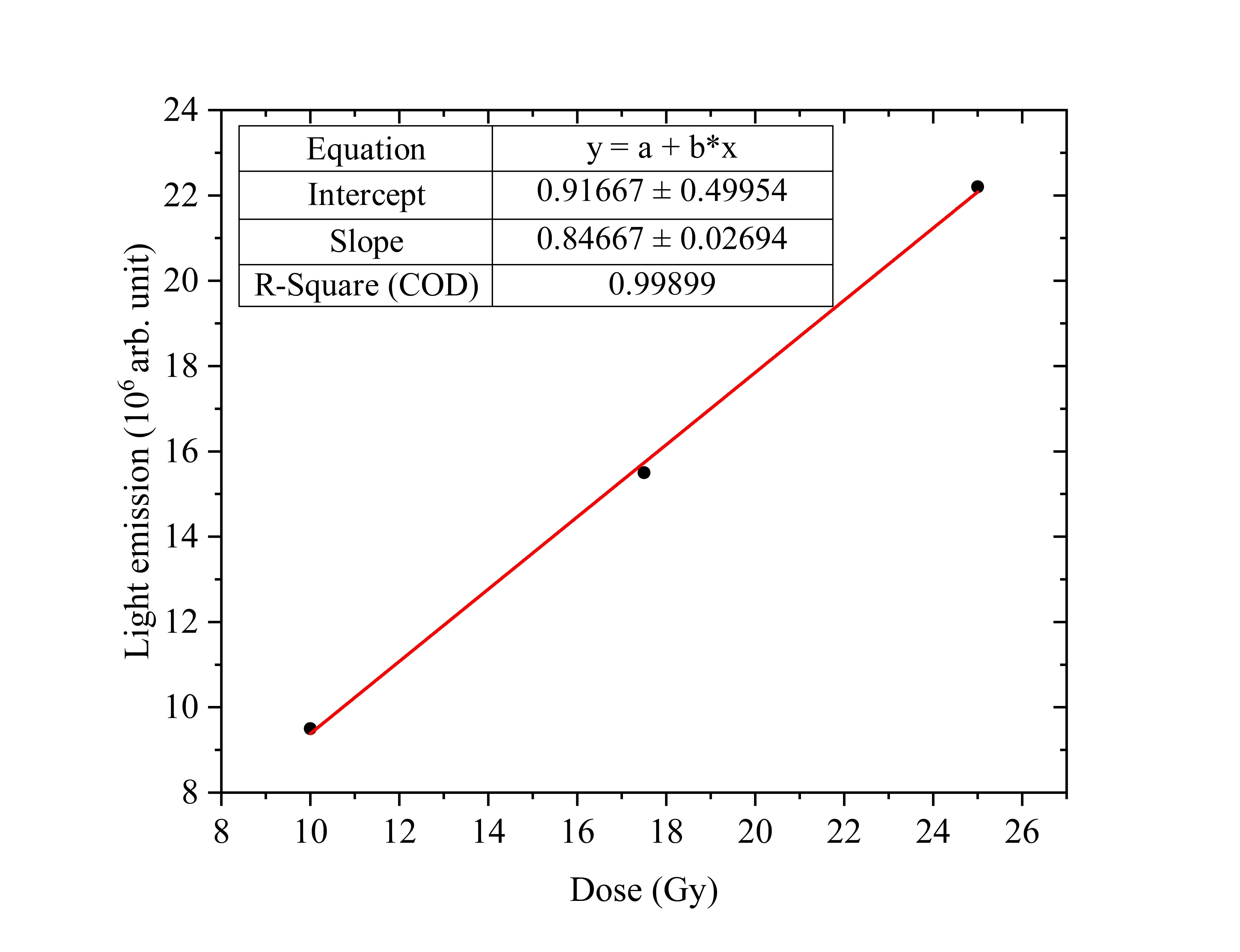}
\caption{The response of the standard CsI(Tl) scintillator to the proton dose.}
\label{fig:3}%
\end{figure}

\indent Figure \ref{fig:2} shows a typical light emission spectrum produced by ionizing a standard CsI(Tl) scintillator with a proton beam. The emitted light spectrum from the CsI(Tl) scintillator was characterized by a dominant single peak centered at 575 nm. To quantify the scintillator's light yield, the integrated area under the emission spectrum was determined within the wavelength range of 350 to 850 nm. The response of the scintillator to the proton beam revealed that the total emitted light was linearly proportional to the total dose of proton irradiation, as indicated in figure \ref{fig:3}. While the scintillator effectively distinguished doses of the proton beam, this relationship was not directly proportional to the total dose delivered by the cyclotron. The size of the scintillator should align with the beam size, enhancing precision in the total dose measurements. The analysis of the dose response suggested that the scintillator was promising to be used as the proton dosimeter for accurately measuring the depth-dose profiles.\\

\begin{figure}[htp] 
\centering
\includegraphics[width=0.8\textwidth]{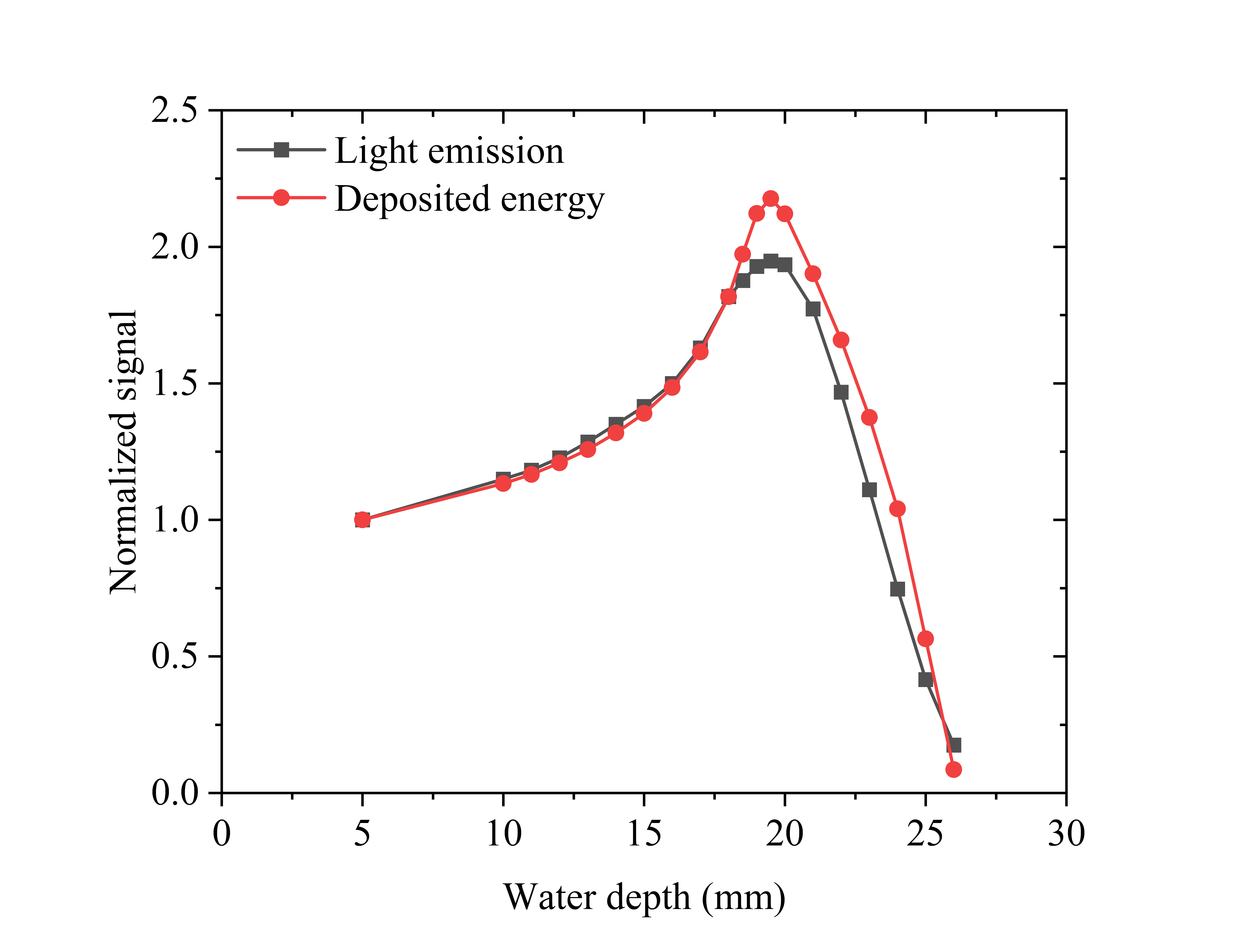}
\caption{Depth-dose profile of the 70 MeV proton energy deposited in the water phantom. }
\label{fig:4}%
\end{figure}

\indent The measured depth-dose profile of the 70 MeV proton beam using the CsI(Tl) scintillator is presented in figure \ref{fig:4}. A comparison between the measured profile and the simulated depth-dose profile revealed good agreement from 5 mm to approximately 19.5 mm of water depth. However, the measured light emission was observed to be lower than the deposited energy, attributed to the well-known quenching phenomenon \cite{ref2,ref9,ref15}. Beyond the quenching region, the measured depth-dose profile of output light showed a decrease compared to the deposited energy of the proton beam from the simulations. Specifically, the peak-to-plateau of the output light at 0.5 cm of water depth was 1.95 in the quenching region, while the simulation exhibited a higher peak-to-plateau ratio of 2.18. A previous study by Almurayshid \cite{ref2} reported a peak-to-plateau of around 2.8 for plastic scintillators at 0.0 cm of water, while measurements by Hoehr \cite{ref9} on Gd$^{3+}$-doped silica fiber yielded a ratio of 3.51, indicating a good agreement between deposited energy and light emission across the depth-dose profile. According to the low ratio in our measurement, the problem was suggested to be resolved by designing a new setup for the measurement of the depth-dose profile.

\begin{figure}[htp] 
\centering
\includegraphics[width=0.8\textwidth]{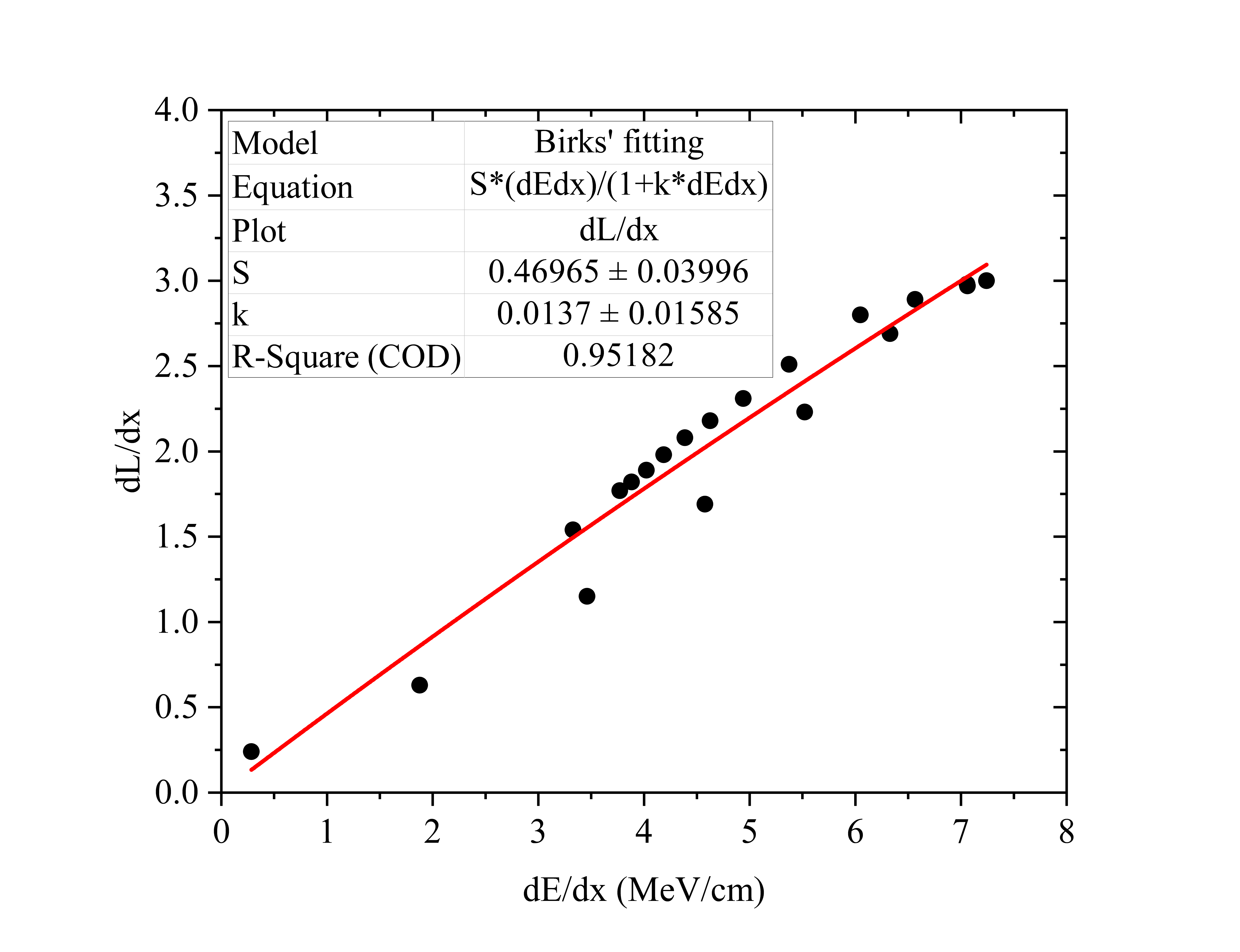}
\caption{Relationship between dE/dx of a 70-MeV proton beam and its corresponding emission dL/dx. The black dots indicate the data sets, which were fitted using equation \ref{eq:brik}. The fitted line is shown in red. The Birks’ constant of the standard CsI(Tl) scintillator used was k$_B$ = 0.0137 $\pm$ 0.0159 cm/MeV.}
\label{fig:5}%
\end{figure}

\indent The measured output light plotted against the deposited energy from the simulations and fitted equation \ref{eq:brik} is shown in figure \ref{fig:5}. The measured output light was the constant of $k_B$ = 0.0137 $\pm$ 0.0159 cm/MeV from the fitting. The curve was fitted into a data range before a quenching region. The Birks’ constant from the results of Torrisi’s measurement \cite{ref4} in the case of a plastic scintillator was around 0.021 cm/MeV. The constant in the measurement of Gd-doped silica fiber \cite{ref9} was 0.016 cm/MeV, and Savard \cite{ref12} presented the Ce$^{3+}$-doped silica fiber with a constant of 0.023 cm/MeV. The standard CsI(Tl) scintillator presented lower quenching by showing the Birks’ constant equivalently to the Gd-doped silica fiber but lower than those from the common plastic scintillator and Ce$^{3+}$-doped silica.

\begin{figure}[htp] 
\centering
\includegraphics[width=0.8\textwidth]{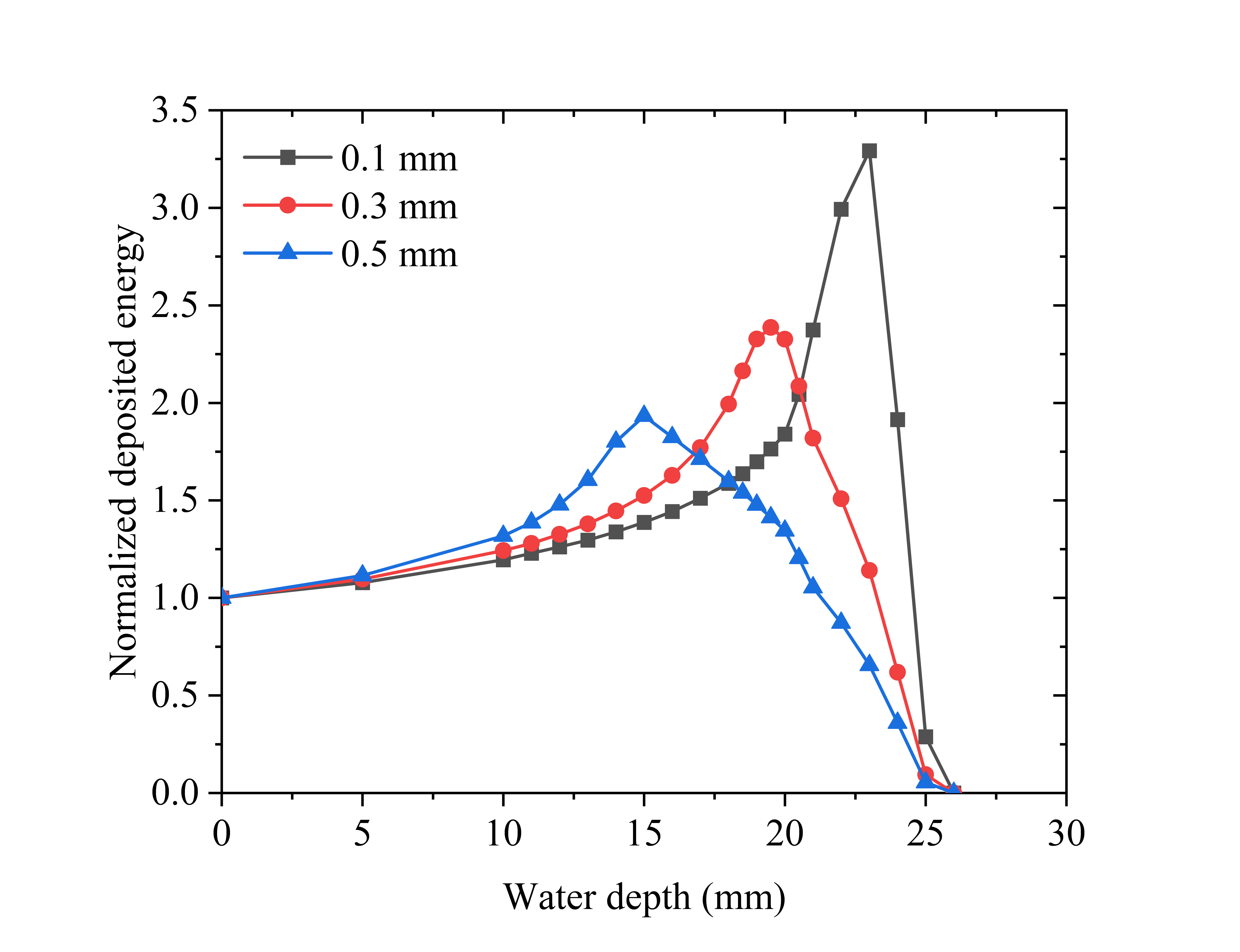}
\caption{The depth-dose of the proton beam at different thicknesses of the scintillator.}
\label{fig:6}%
\end{figure}

\indent To achieve the raw depth-dose profile, the simulations were calculated by modifying the size of the scintillator as well as the thickness of the water tank. The size of the scintillator was chosen based on its thickness, which was 0.1, 0.3, and 0.5 cm. Figure \ref{fig:6} presents the characteristic depth-dose for each thickness of the scintillator. The thin scintillator showed a narrow Bragg peak and a deeper Bragg peak. In contrast, the thicker scintillator showed a shallow Bragg peak and a broader Bragg peak because it received the deposited energy over a longer range compared to the thinner thickness. However, the last point of the energy deposition for each thickness was the same because the proton beam completely stopped within the scintillator. Therefore, the shape of the depth-dose profile due to deposited protons energy into the scintillator depended on the scintillator thickness. These results were consistent with Archambault’s work \cite{ref14}, which found that the larger thickness of the scintillator provided a broader Bragg peak. In figure \ref{fig:6}, the simulated peak-to-plateau values (at 0.0 cm of water) were 3.29, 2.39, and 1.93 for the scintillator thickness of 0.1, 0.3, and 0.5 cm, respectively. The peak-to-plateau values of the 0.1 cm-thickness scintillator matched closely with the measurements reported in \cite{ref9}. Consequently, reducing the thickness of the scintillator leads to increased peak-to-plateau values, resulting in a higher accuracy of the depth-dose profile.

\begin{figure}[htp] 
\centering
\includegraphics[width=0.8\textwidth]{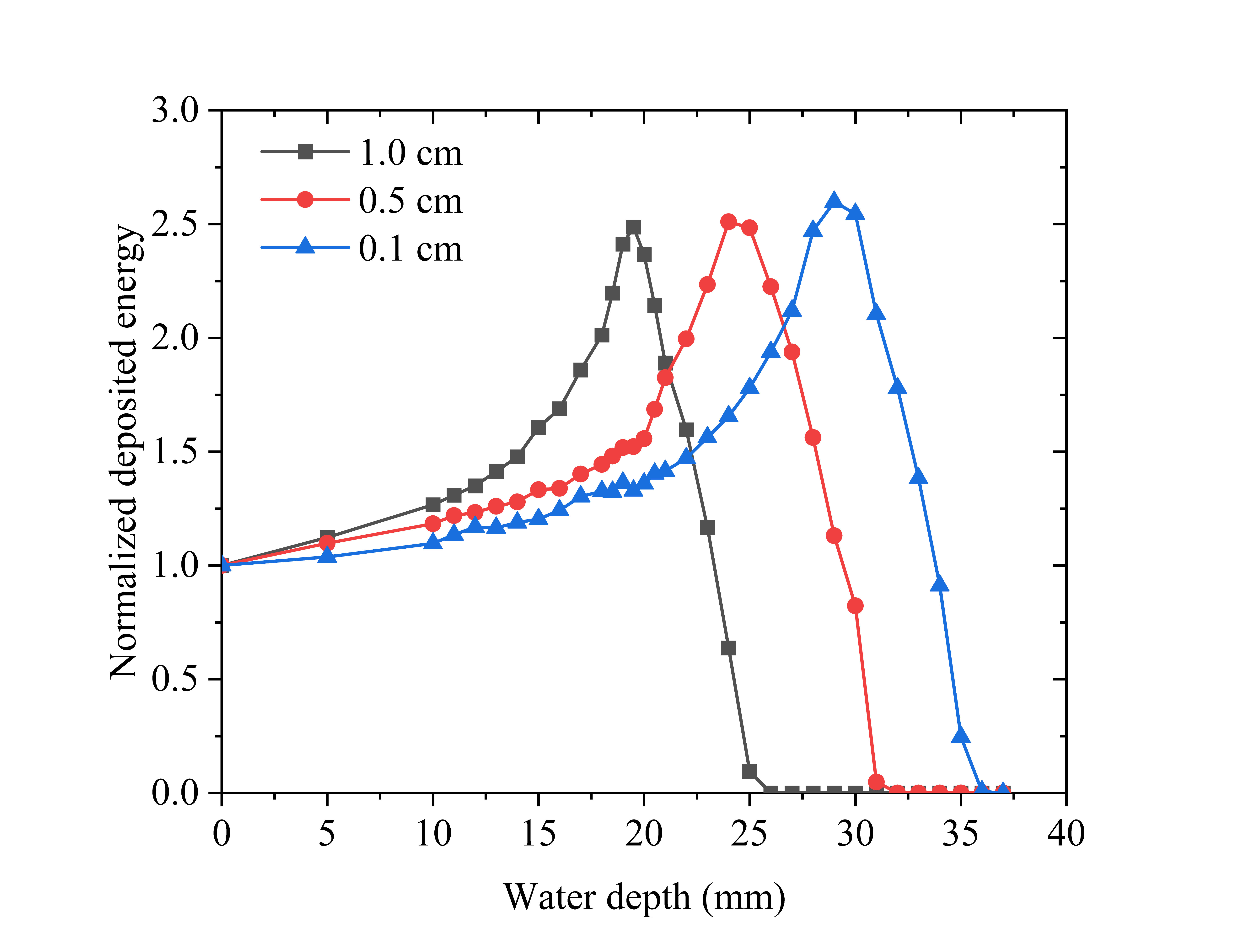}
\caption{The depth-dose profile of the proton beam at different thicknesses of the water tank. The deposited energy for each tank thickness was normalized to the highest deposited energy.}
\label{fig:7}%
\end{figure}

\indent The simulation results of varying the tank thickness are shown in figure \ref{fig:7}. The narrower tank wall's thickness showed that Bragg peak regions shifted deeper compared to the thicker wall. For the initial energy of the proton beam after passing through the tank wall, the average energies of tracked protons in the scintillator are indicated in Table \ref{tab:1}. The average energy of tracked protons inside the scintillator decreased when the water tank thickness increased. The decrease in protons occurred because the remaining protons scattered throughout the water tank \cite{ref1}. The 0.1-cm water tank had a higher measured peak-to-plateau (at 0.0 cm of water) compared to other thicknesses due to the higher energy deposition of the protons. Thus, the simulations of varying the water tank thickness revealed that the water tank thickness acted as a range shifter by causing protons to lose energy to the water tank, which was agree to \cite{ref9}.

\begin{table}[htp]
\centering
\caption{The average energy of the tracked proton inside the scintillator for each water tank thickness. The remaining protons were compared to those of the setup without the water tank.}
\label{tab:1}%
\begin{tabular}{ccc}

\hline
Water tank thickness (cm) & Average energy (MeV) & \% Remaining protons \\ \hline
0.1                       & 65.69 $\pm$ 1.55     & 99.54                      \\
0.5                       & 60.85 $\pm$ 1.81     & 98.82                     \\
1.0                         & 54.39 $\pm$ 1.78     & 97.89                     \\ \hline
\end{tabular}
\end{table}

\indent This work proposes the proper design for measuring the depth-dose profile of the proton beam. To further receive the raw depth-dose profile, the scintillator with a thickness of 0.1 cm or less was recommended for determining the depth-dose profile of the proton beam \cite{ref8,ref9,ref14}. Moreover, the scintillator should be coated with cladding to protect the background light and enhance light transmittance \cite{ref20,ref21}. Despite functioning as a range shifter for measuring the horizontal movement of the water tank, it is advisable to minimize the wall thickness of the tank to acquire the raw depth-dose profile \cite{ref9,ref22}. The thickness of the water tank not only functions as a range shifter but can also be employed to measure the depth-dose profile.

\section{Conclusions}
\indent The depth-dose profile of the proton beam was measured using the standard CsI(Tl) scintillator and a water phantom. The scintillator can distinguish the proton dose, which depends on the quantity of the light yield. Our setup, consisting of a 0.3-cm scintillator and a 1.0-cm water tank, was able to measure the depth-dose profile. The shape of the depth-dose profile of the 70-MeV proton beam was in good agreement with the deposited energy from the simulation until the quenching region. Our setup presented that the Birks' constant from the fitting was $k_B$ = 0.0137 $\pm$ 0.0159 cm/MeV. The depth-dose measurement suggested that the 0.1 cm scintillator produced a narrower Bragg peak compared to thicker scintillators. The thickness of the tank wall acted as a range shifter, reducing the initial energy of the incoming protons. This study demonstrated that the standard CsI(Tl) scintillator is an effective tool for measuring proton doses and depth-dose profiles, making it a viable option for detecting proton beam characteristics.

\acknowledgments
\indent This project was supported by (i) Ubon Ratchathani University (UBU), (ii) Suranaree University of Technology (SUT), (iii) Thailand Institute of Nuclear Technology (TINT), and (iv) King Chulalongkorn Memorial Hospital (KCMH) to facilitate laboratory work. We would like to acknowledge funding  from the National Science, Research, and Innovation Fund (NSRF), Project No. FF66-B37G660013, and the Thailand Center of Excellence in Physics (ThEP), Project No. ThEP-61-PHM-SUT4.

\end{document}